\documentclass{elsart}
\usepackage{amsmath}
\usepackage{amsfonts}
\usepackage{amssymb}
\usepackage{graphics}
\usepackage{latexsym}
\usepackage{euscript}
\usepackage{supertabular}
\usepackage[spanish, american]{babel}

\begin{document}
\selectlanguage{american}
\numberwithin{equation}{section}
\numberwithin{table}{section}

%\input epsf
%%% Blackboard bold "1". Not in the AMS font set.
\def\Bid{{\mathchoice {\rm {1\mskip-4.5mu l}} {\rm
{1\mskip-4.5mu l}} {\rm {1\mskip-3.8mu l}} {\rm {1\mskip-4.3mu l}}}}
%%%

\newcommand{\eL}{{\cal L}}
\newcommand{\J}{\textbf{J}}
\newcommand{\bP}{\textbf{P}}
\newcommand{\G}{\textbf{G}}
\newcommand{\K}{\textbf{K}}
\newcommand{\M}{{\cal M}}
\newcommand{\E}{{\cal E}}
\newcommand{\bu}{\textbf{u}}
\newcommand{\tr}{\mbox{tr}}
\newcommand{\norm}[1]{\left\Vert#1\right\Vert}
\newcommand{\abs}[1]{\left\vert#1\right\vert}
\newcommand{\set}[1]{\left\{#1\right\}}
\newcommand{\ket}[1]{\left\vert#1\right\rangle}
\newcommand{\bra}[1]{\left\langle#1\right\vert}
\newcommand{\ele}[3]{\left\langle#1\left\vert#2\right\vert#3\right\rangle}
\newcommand{\inn}[2]{\left\langle#1\vert#2\right \rangle}
\newcommand{\Real}{\Bid R}
\newcommand{\dmat}[2]{\ket{#1}\!\!\bra{#2}}

\begin{frontmatter}

\title{Volumes of Compact Manifolds}

\selectlanguage{spanish}
\author{Luis J. Boya}
\address{Departmento de F\'isica Te\'orica\\
Facultad de Ciencias\\
Universidad de Zaragoza\\
E-50009 Zaragoza, Spain}
\ead{luisjo@posta.unizar.es}

\noextrasspanish
\selectlanguage{american}

\author{E.C.G. Sudarshan}
\address{Center for Particle Physics \\
Physics Department \\
The University of Texas at Austin \\
Austin, Texas 78712-1081}
\ead{sudarshan@physics.utexas.edu}

\author{Todd Tilma}
\address{Digital Materials Lab \\
Frontier Research System \\
The Institute of Physical and Chemical Research\\
Wako-shi, Saitama-ken, 351-0198 Japan}
\ead{tilma@riken.jp}

\date{\today}

\begin{abstract}
We present a systematic calculation of the volumes of \textit{compact}
manifolds which appear in physics: spheres, projective spaces, group manifolds and 
generalized flag manifolds.  In each case we state what we believe is the most
natural scale or normalization of the manifold, that is, the generalization of the 
unit radius condition for spheres.  For this aim we first describe the manifold
with some parameters, set up a metric, which induces a volume element, and
perform the integration for the adequate range of the parameters; in most cases our manifolds will
be either spheres or (twisted) products of spheres, or quotients of spheres
(homogeneous spaces).

Our results should be useful in several physical instances, as instanton 
calculations, propagators in curved spaces, sigma models, geometric scattering
in homogeneous manifolds, density matrices for entangled states, etc.  Some flag manifolds
have also appeared recently as exceptional holonomy manifolds; the volumes of
compact Einstein manifolds appear in String theory.
\end{abstract}

%\flushleft{03.65.Bz, 03.65.-w, 03.67.L}

\end{frontmatter}

%------------------------------------------------------------------------
%------------------------------------------------------------------------
\section{Introduction}
\label{sec:intro}
Interest in volumes of compact manifolds appearing in physics arises for several
reasons.  For example, in instanton calculations it is necessary to integrate over the 
gauge group \cite{tHoft}; in quantum field theories over curved spaces, the
volumes of the compact spaces enter very often \cite{deWitt}.  For entangled qubit states
the geometry of the corresponding groups $SU(2n)$ and quotient spaces $SU(2n)/U(n)\times U(n)$
are important; etc.

The main purpose of this paper is to compute the volumes of several types of compact
manifolds in a systematic way.  We proceed from the most symmetric cases
(spheres) to the less symmetric (flag manifolds).  Essential
ingredients are the scale (``radius'') of the manifold, the setting of coordinates and their
ranges.  All these spaces will be Riemannian, so we compute the
volume with respect to that metric; all our manifolds will also be
\textit{homogeneous} manifolds $\EuScript{M} = G/K$.

We define the manifold $\EuScript{M}$ in terms of some parameters $\{\alpha^i\}$
in some local charts, so as to cover the whole manifold.  The metric will look
locally as
\begin{equation}
\label{oneone}
g = g_{ij}(\alpha_k)d\alpha^i d\alpha^j;
\end{equation}
the associated volume element is
\begin{equation}
\label{onetwo}
\tau = \sqrt{Det \{g_{ij}\} }\; d\alpha^1 \wedge \cdots \wedge d\alpha^n
\end{equation}
and the volume is (for one chart only)
\begin{equation}
\label{onethree}
\text{Vol}(\EuScript{M}) = \int \cdots \int_{\alpha_{0}^i}^{\alpha_{1}^i} \cdots \int \tau
\end{equation}
where $\alpha_{0}^i, \alpha_{1}^i$ is the range of the $i^{th}$ parameter;
much of our discussion will be about these ranges.

In all cases the scale of the object to be measured will be stated in a ``natural'' manner;
we shall make comparison with other choices made in the literature when convenient:
it is not always easy to recognize the different conventions used.  In fact, it is only for spheres
that the universal criterion is the unit radius, but we shall see that even in this case
when we have a manifold described as a \textit{product} of spheres, the volume will \textit{not} always 
be the product of the natural sphere volumes!

This volume problem has been dealt with in the published literature, both physical and
mathematical.  A fairly complete treatment for group manifolds is presented 
by Marinov \cite{Marinov}, corrected in \cite{Marinov2}; the volume of the group
is extracted from the propagator by an indirect (and complicated) method, in fact as a 
by-product of the author's study of curved path integrals \cite{Marinov3}.  In
MacDonald \cite{MacDonald} there is a general synthetic formula for any compact
Lie group, with arbitrary scale, in terms of the root lattice; we feel that most
physicists will find these formulas hard to apply.  For the unitary unimodular group
$SU(n)$ there is the very clear paper by C.\ Bernard \cite{Bernard}.

There are many other partial results in the physics literature, some of them contradictory.  It
was in an effort to overcome these difficulties that we undertook this investigation.
A direct antecedent of our paper is the work of M.\ Byrd \cite{MByrdp1,MByrdp2}, who
studied the $SU(3)$ case thoroughly.  In a further paper we intend to look at the 
geometry of entangled states more closely \cite{Collab1}; see also the papers 
\cite{Tilma1,Tilma2}.

We start by collecting some formulas for \textit{spheres}, as the most symmetric spaces:
the isometry group (in physics, this is called the little group) acts 
transitively in the bundle of bi-planes, and therefore the
\textit{sectional} curvature is constant.  Next is the case of \textit{projective} spaces, in which 
the isometry group still acts transitively in the bundle of lines, and therefore the
geodesics are all equal.  Compact \textit{group manifolds} are symmetric spaces and therefore the
curvature is covariantly constant.  There is a natural metric (induced from the
Killing form in the Lie algebra) which generates an invariant measure (Haar, or more
appropriately Hurwitz), but also there is a structure close to the product of odd spheres:
the two criteria clash.  The action of the group in the adjoint representation produces
interesting orbits, the so-called \textit{generalized flag manifolds}, which appear very often
in geometric quantization, density matrices, entangled states, etc.  We compute also
the volumes of these quotient manifolds; there are still symmetric, or at least homogeneous
manifolds, and therefore the \textit{scalar} curvature is still constant \cite{Wol}.
%---
%---
\section{Spheres}
\label{sec:two}
The sphere $S^{d-1}$ is a maximally symmetric space; in fact, the isometry group $O(d)$
is transitive (so spheres are homogeneous spaces) and the isotropy group
$O(d-1)$ acts also \textit{trans} (from here on, we shall use \textit{trans} for ``transitive'') 
in the unit tangent vectors 
(hence permutes the geodesics, and all are equivalent, in particular closed with same length)
and also \textit{trans} in the Grassmannian of bi-planes (hence the sectional
curvature is constant, in particular the scalar curvature). We have
\begin{align}
S^{d-1} =& \frac{O(d)}{O(d-1)}, \nonumber \\
X \equiv \mathbb{R}\mbox{P}^{d-2} = \frac{O(d-1)}{O(1)\times O(d-2)}, &\quad
Y \equiv \text{\textit{Gr}}_{d-1,2} = \frac{O(d-1)}{O(2)\times O(d-3)}
\label{eq:twoone}
\end{align}
where $X$ is the space of lines and $Y$ that of bi-planes, in the tangent space
to a point.

The volume of the sphere of \textbf{unit radius} embedded in $\mathbb{R}^d$, $d\geq 1$, is
calculated from the auxiliary formula
\begin{equation}
\label{twotwo}
[\int_{-\infty}^{+\infty} e^{-x^2}dx ]^n = \pi^{\frac{n}{2}} = \int_{0}^{\infty} r^{n-1} e^{-r^2} dr
\times \text{Vol}(S^{n-1})
\end{equation}
or
\begin{equation}
\label{twothree}
\text{Vol}(S^{d-1}) = \frac{2\pi^{\frac{d}{2}}}{\Gamma(\frac{d}{2})}.
\end{equation}
For completeness we include a proof by induction.  If $S^{d-1}$ is
embedded in $S^d$ as the equator, $S^{d-1} \subset S^d$ and $\theta_d$ is the latitude
angle, with $0 \leq \theta_d \leq \pi$, we have for the metric
\begin{equation}
\label{twofour}
g(S^{d}) = ds_d^2 = d\theta_d^2 + \sin^2(\theta_d) ds_{d-1}^2.
\end{equation}
We start by setting
\begin{gather}
\text{Vol}(S^0)= \sharp \{\text{North, South}\} = 2 \nonumber \\
\quad 0 \leq \theta_i \leq \pi \quad \text{for i: }1, 2, \ldots, d.
\label{eq:twofive}
\end{gather}
Therefore as
\begin{equation}
\label{twoseven}
dV_{S^d} = \tau(S^d) = \sqrt{Det\{g\}} \; \prod_{i=1}^d d\theta_i \cdot 2
\end{equation}
we find
\begin{align}
\text{Vol}(S^d) &= \int_{0}^\pi \sin^{d-1}(\theta_d) \int_{0}^\pi \sin^{d-2}(\theta_{d-1}) \cdots
\int_{0}^\pi d\theta_1 \cdot 2 \nonumber \\
&= \int_{0}^\pi \sin^{d-1}(\theta_d) d\theta_d \times 
\text{Vol}(S^{d-1}).
\label{eq:twoeight}
\end{align}
In particular
\begin{align}
\text{Sphere: }&S^0,\;S^1,\;S^2,\;S^3,\;S^4,\;S^5 \nonumber \\
\text{Vol}(S^d):\; &2,\;2\pi,\;4\pi,\;2\pi^2,\;\frac{8\pi^2}{3},\;\pi^3.
\label{eq:twonine}
\end{align}
Explicit useful formulas for the even/odd cases are
\begin{equation}
\label{twoten}
\text{Vol}(S^{2n+1}) = \frac{2\pi^{n+1}}{n!}
\quad \text{ and } \quad
\text{Vol}(S^{2n}) = \frac{2(2\pi)^n}{(2n-1)!!}.
\end{equation}
Notice the volume is maximal for $S^6$, and 
\begin{equation}
\label{twotwelve}
\underset{n\rightarrow \infty}{lim}\; \text{Vol}(S^n) \rightarrow 0.
\end{equation}
These ``volumes'' are numbers, they have no dimensions (in hypersolid angles).  
Our convention for the radius $R=1$ is \textit{extrinsic}, that is, 
depends on the embedding $S^{d-1} \subset \mathbb{R}^d$;
an equivalent \textit{intrinsic} criterion is the length of the geodesic:
\begin{equation}
\label{twothirteen}
\text{geodesic length} = 2\pi \leftrightarrow \text{ radius }R =1 
\end{equation}
applicable because all geodesics in the sphere are equivalent (see remarks
above).  Notice also that the scalar curvature is
\begin{equation}
\label{twofourteen}
R_{sc}(S_R^n) = \text{Tr}[Ricci] = \text{Tr}[\text{Tr}[Riemann]] = \frac{n(n-1)}{R^2}
\end{equation}
(see \cite{Geom1,Hel1}), so $R_{sc}$ is not a good scale as it depends on $n$.  Finally the volume of the unit ball
is
\begin{equation}
\label{twofifteen}
\text{Vol}(D^n) = \int_{0}^1 r^{n-1}(\text{Vol}(S^{n-1}))dr = \frac{\text{Vol}(S^{n-1})}{n}.
\end{equation}
%---
%---
\section{Projective Spaces}
\label{sec:three}
The projective spaces, $\mathbb{R}\mbox{P}^n$, $\mathbb{C}\mbox{P}^n$, $\mathbb{H}\mbox{P}^n$, and
$\mathbb{O}\mbox{P}^2$ share with the spheres the property of being two-point homogeneous spaces, or
symmetric rank-one spaces \cite{Hel1}: the isometry (little) groups act \textit{trans} on directions,
and hence permute the geodesics, which (in the compact case) would all have to be closed
and of the same length.  These spaces are of constant \textit{covariant} curvature, as they are
symmetric, but not of constant \textit{sectional} curvature, except $\mathbb{R}\mbox{P}^n$ of course.  
One can see how the sectional curvature
changes along the real bi-planes e.\ g.\ in $\mathbb{C}\mbox{P}^2$ \cite{Santan}.

So we have a natural characterization for the scale of projective spaces: namely, the length
of a fiducial geodesic. We shall define projective spaces as \textit{quotient of spheres}, and we shall see
that the natural scale for the geodesics is to have length $\pi$, \textit{not} $2\pi$ as in
the spheres.
%---
\subsection{Real Projective Spaces}

$\mathbb{R}\mbox{P}^n$ is defined as the set of rays or lines or one-dimensional sub-spaces in 
$\mathbb{R}^{n+1}$:
\begin{equation}
\label{threeone}
\mathbb{R}\mbox{P}^n = \{\text{lines in $\mathbb{R}^{n+1}$}\}.
\end{equation}
There are many equivalent characterizations:
\begin{equation}
\label{threetwo}
\mathbb{R}\mbox{P}^n = \frac{S^n}{S^0} \equiv \frac{S^n}{\{\pm1\}} \equiv \frac{S^n}{Z_2} 
\equiv \frac{S^n}{\text{anti-podal map}} = \frac{O(n+1)}{O(n)\times O(1)} = \frac{SO(n+1)}{O(n)}.
\end{equation}
Any line touches the unit sphere in $\mathbb{R}^{n+1}$ in the two anti-podal points, hence
the first characterizations.  The orthogonal group $O(n+1)$ acts \textit{trans} in the lines,
with little group: stabilizer of a vector in $O(n)$ times the two directions in $O(2)$, hence the last
forms in equation \eqref{threetwo}.

Now we \textit{define} the volume of $\mathbb{R}\mbox{P}^n$ as a quotient:
\begin{equation}
\label{threethree}
\text{Vol}(\mathbb{R}\mbox{P}^n)= \frac{\text{Vol}(S^n)}{\text{Vol}(S^0)} = \frac{\text{Vol}(S^n)}{2} 
= \frac{\pi^{\frac{n+1}{2}}}{\Gamma(\frac{n+1}{2})}.
\end{equation}
But this implies that the geodesic length of \textit{our} $\mathbb{R}\mbox{P}^n$ is $\pi$ and not
$2\pi$ because the half-meridian from the the North pole to the South pole is already \textit{closed}
in $\mathbb{R}\mbox{P}^n$!  Notice the halving of the volume for $\mathbb{R}\mbox{P}^n$ is really neglecting
a set of measure zero, because $\mathbb{R}\mbox{P}^n$ as a half-sphere, still has the anti-podal points
in the equator identified; but this is a set of measure zero.
Thus we get $\text{Vol}(\mathbb{R}\mbox{P}^1)=\pi$, even though $\mathbb{R}\mbox{P}^1 \cong S^1$ and
$\text{Vol}(S^1) = 2\pi$: the radius has shrunk to 1/2.  The case $n=3$ is also interesting,
because $\mathbb{R}\mbox{P}^3 = SO(3)$, and we shall discuss it later.
Notice also $\mathbb{R}\mbox{P}^{2n}$ is not orientable, hence the volume has to be properly defined with the
modulus of the measure.

For $CROSS$ spaces (the notation, due to Besse \cite{Besse1}, means $\underline{C}$ompact, $\underline{R}$ank-$\underline{O}$ne,
$\underline{S}$ymmetric, $\underline{S}$paces, precisely the sphere and the projective spaces)
A.\ Weinstein has established the following result \cite{Wei1}, which we shall not prove but verify in many examples.
If $\EuScript{M}$ is a $CROSS$,
dim $\EuScript{M}=n$, geodesic length=$l$, then
\begin{equation}
\label{threefour}
\biggr(\frac{2\pi}{l}\biggl)^n\frac{\text{Vol}(CROSS)}{\text{Vol(Sphere same dimension)}} = \textit{integer} \equiv i(CROSS)
\end{equation}
with equal normalization.  For our case of real projective spaces we indeed get
\begin{equation}
\label{threefive}
i(\mathbb{R}\mbox{P}^n) = \frac{\text{Vol}(\mathbb{R}\mbox{P}^n)}{\text{Vol}(S^n)}\biggr({\frac{2\pi}{\pi}}\biggl)^n = \frac{1}{2}\;\cdot 2^n = 2^{n-1}
\end{equation} 
in agreement with Besse \cite{Besse1}.
%---
\subsection{Complex Projective Spaces}

$\mathbb{C}\mbox{P}^n$ has the following definition:
\begin{align}
\mathbb{C}\mbox{P}^n = \{\text{Set of lines in $\mathbb{C}^{n+1}$}\} &= \frac{S^{2n+1}}{S^1} = \frac{S^{2n+1}}{U(1)} \nonumber \\
&= \frac{U(n+1)}{U(1)\times U(n)} =\frac{SU(n+1)}{U(n)}.
\label{eq:threesix}
\end{align}
Here the complex lines in $\mathbb{C}^{n+1}$ intersect the unit sphere $S^{2n+1}$ in $\mathbb{C}^{n+1} \equiv 
\mathbb{R}^{2n+2}$ along a maximal circle $S^1$.  We again \textit{define} the volume of $\mathbb{C}\mbox{P}^n$ 
as a quotient
\begin{equation}
\label{threeseven}
\text{Vol}(\mathbb{C}\mbox{P}^n) = \frac{\text{Vol}(S^{2n+1})}{\text{Vol}(S^1)} = \frac{2\pi^{n+1}/n!}{2\pi} = \frac{\pi^n}{n!}
\end{equation}
using equation \eqref{twoten}.  Again for $n=1$ we have $\text{Vol}(\mathbb{C}\mbox{P}^1) = \pi$, whereas 
$\mathbb{C}\mbox{P}^1 = S^3/S^1 \cong S^2$, and $\text{Vol}(S^2) = 4\pi$: obviously the geodesic length of \textit{our}
$\mathbb{C}\mbox{P}^1$ is only $\pi$, whereas the volume of the equivalent $S^2$ space is $4 = 2 \cdot 2$ times as big: the
quotient circle includes the anti-pode ($S^0$ lies inside $S^1$),
hence the geodesics are halved.  It is remarkable that
\begin{equation}
\sum_{n=0}^{\infty} \text{Vol}(\mathbb{C}\mbox{P}^n) = e^{\pi} \approx 23.147.
\end{equation}

The Weinstein integer for $\mathbb{C}\mbox{P}^n$ is also easy to compute:
\begin{align}
i(\mathbb{C}\mbox{P}^n) &= \frac{\text{Vol}(\mathbb{C}\mbox{P}^n)}{\text{Vol}(S^{2n})}\biggr(\frac{2\pi}{\pi}\biggl)^{2n} 
= \frac{\pi^n/n! \cdot 2^{2n}}{2(2\pi)^n/(2n-1)!!} = \frac{(2n-1)!!}{n!} \; 2^{n-1} \nonumber \\
&= \frac{(2n-1)!}{(n-1)!n!} = \binom{2n-1}{n-1}
\label{eq:threeeight}
\end{align}
in full agreement with Besse \cite{Besse1}.  For a thorough mathematical study of $\mathbb{C}\mbox{P}^2$ as an instanton see \cite{Gib}.
%---
\subsection{Quaternionic Projective Spaces}

$\mathbb{H}\mbox{P}^n$ has the following definition: let $\mathbb{H}$ be the space of quaternions and $S^3$ the set of unit quaternions
and $Sp(n)$, the symplectic group $C_n$.
\begin{equation}
\label{threenine}
\mathbb{H}\mbox{P}^n = \{\text{lines in $\mathbb{H}^{n+1} \cong \mathbb{R}^{4n+4}$}\} = \frac{S^{4n+3}}{S^3}  = \frac{S^{4n+3}}{Sp(1)}
=\frac{Sp(n+1)}{Sp(n) \times Sp(1)}.
\end{equation}
With the same definition as for $\mathbb{K} = \mathbb{R}$ or $\mathbb{C}$, we obtain
\begin{equation}
\label{threeten}
\text{Vol}(\mathbb{H}\mbox{P}^n) = \frac{\text{Vol}(S^{4n+3})}{\text{Vol}(S^3)} = \frac{2\pi^{2n+2}/(2n+1)!}{2\pi^2}
=\frac{\pi^{2n}}{(2n+1)!}
\end{equation}
with \textit{geodesic length} equal to $\pi$.  Again $S^3$ contains the anti-podal point.  We obtain 
$\text{Vol}(\mathbb{H}\mbox{P}^1) = \pi^2/6$, whereas $\mathbb{H}\mbox{P}^1 \cong S^4$ and $\text{Vol}(S^4) = 8\pi^2/3$.
In this case 
\begin{equation}
\sum_{n=0}^{\infty} \text{Vol}(\mathbb{H}\mbox{P}^n) = \frac{\sinh(\pi)}{\pi} =
\frac{\text{\textit{I}}_{\frac{1}{2}}(\pi)}{\sqrt{2}} \approx 3.676,
\end{equation}
where $\text{\textit{I}}_{\frac{1}{2}}(\pi)$ is the modified Bessel
function of the first kind.  

The Weinstein number is
\begin{align}
i(\mathbb{H}\mbox{P}^n) &= \frac{\text{Vol}(\mathbb{H}\mbox{P}^n)}{\text{Vol}(S^{4n})}\biggr(\frac{2\pi}{\pi}\biggr)^{4n} = 
\frac{\pi^{2n}/(2n+1)!}{2(2\pi)^{2n}/(4n-1)!!}\; 2^{4n} \nonumber \\
&= \frac{(4n-1)!!}{(2n+1)!}\;2^{2n-1}
= \frac{(4n-1)!}{(2n+1)(2n)!(2n-1)!} \nonumber \\
&= \frac{1}{2n+1}\binom{4n-1}{2n-1}
\label{eq:threeeleven}
\end{align}
in full agreement with Besse \cite{Besse1}.
%---
\subsection{Octonion Projective Space}

Over the division algebra of the octonions $\mathbb{O}$ (or Cayley numbers) there are only $\mathbb{O}\mbox{P}^1$,
which is equal to $S^8$ and $\mathbb{O}\mbox{P}^2$ (the Moufang plane), due to the non-associativity of the 
octonions; and in fact, the definition is different.  For example, $\mathbb{O}\mbox{P}^2$ is best defined as
the hermitian, idempotent, trace-one elements of the exceptional 3\;x\;3 Jordan algebra over
the octonions \cite{Hel1,Gilmore,Rosen}.  But the formulas work just as well and one can formally define:
\begin{equation}
\label{threetwelve}
\text{``}\mathbb{O}\mbox{P}^n\text{''} = \{\text{lines in $\mathbb{O}^{n+1} \cong \mathbb{R}^{8n+8}$}\} = \frac{S^{8n+7}}{S^7}.
\end{equation}
In fact, the only two cases are $\mathbb{O}\mbox{P}^1=S^{15}/S^7 = Spin(9)/Spin(8)$ and $\mathbb{O}\mbox{P}^2 = F_4/Spin(9)$
\begin{align}
\delta :\; S^7 \rightarrow S^{15} \rightarrow S^{8} &= \mathbb{O}\mbox{P}^1 \nonumber \\
Spin(9) \rightarrow F_4 &\rightarrow \mathbb{O}\mbox{P}^2
\label{eq:threethirteen}
\end{align}
where $\delta$ is the fourth Hopf bundle \cite{Geom1}, which is not principal; for $\mathbb{O}\mbox{P}^2$ see 
\cite{Hel1,Rosen}.  $F_4$ is the fifty-two dimensional second exceptional Lie group.  Thus
\begin{align}
\frac{\text{Vol}(S^{8n+7})}{\text{Vol}(S^7)} &= \frac{3!\pi^{4n}}{(4n+3)!}, \nonumber \\
\text{Vol}(\mathbb{O}\mbox{P}^1) &= \frac{\pi^4}{7\cdot 6\cdot 5\cdot 4} = \frac{1}{2^8}
\biggr(\text{Vol}(S^8) = \frac{32\pi^4}{7\cdot 5\cdot 3}\biggl),
\label{eq:threefourteen}
\end{align}
and
\begin{equation}
\label{threefifteen}
\text{Vol}(\mathbb{O}\mbox{P}^2) = \frac{3!\pi^8}{11!}.
\end{equation}
We have Weinstein integer $i(\mathbb{O}\mbox{P}^1)$=1 obviously, and
\begin{equation}
\label{threesixteen}
i(\mathbb{O}\mbox{P}^2) = \frac{\text{Vol}(\mathbb{O}\mbox{P}^2)}{\text{Vol}(S^{16})}\; 2^{16} = 3 \cdot 13 = 39
\end{equation}
as it should \cite{Besse1}.

We have grouped the main results from sections \ref{sec:two} and \ref{sec:three} in Table \ref{tbl:A}:
\tablehead{ \multicolumn{5}{c}{\textbf{Table \ref{tbl:A}: Volumes of Spheres and Projective Spaces}}\\\hline
Manifold & Symbol & Normalization & Volume & Notes \\\hline \hline}
\begin{center}
\begin{supertabular}{|l|c|l|c|l|}
\label{tbl:A}
Spheres & $S^{d-1}$ & radius, $R=1$ & $\frac{2\pi^{\frac{d}{2}}}{\Gamma(\frac{d}{2})}$
& Maximally symmetric \\
 & & geodesic length = $2\pi$ & & \\\hline
Real Projective & $\mathbb{R}\mbox{P}^n$ & geodesic length = $\pi$ & $\frac{\pi^{\frac{n+1}{2}}}{\Gamma(\frac{n+1}{2})}$
& Compact, Rank One, \\
& & & & Symmetric Space \\ 
& & & & [$CROSS$] \\\hline
Complex  Projective& $\mathbb{C}\mbox{P}^n$ & geodesic length = $\pi$ & $\frac{\pi^n}{n!}$ & [$CROSS$] \\\hline
%Projective & & & & \\\hline
Quaternionic Projective & $\mathbb{H}\mbox{P}^n$ & geodesic length = $\pi$ &$\frac{\pi^{2n}}{(2n+1)!}$ 
& [$CROSS$] \\\hline
%Projective & & & & \\\hline
Moufang Plane& $\mathbb{O}\mbox{P}^2$ & geodesic length = $\pi$ & $\frac{3!\pi^8}{11!}$
& [$CROSS$] \\\hline
%Plane & & & & \\\hline
\end{supertabular}
\end{center}

%---
%---
\section{Group Manifolds: $SU(2)$ and $SO(3)$}
\label{sec:four}
Group manifolds are still symmetric spaces (i.\ e.\ with covariant constant curvature) of rank
$\geq$ 1.  We start with the $r=1$ case which includes $SU(2)$ and $SO(3)=SU(2)/Z_2$ which will reveal already some
complications.

For $G=SU(2)$:
\begin{equation}
\label{fourone}
SU(2) \ni u = 
\begin{pmatrix}
z_{1} & z_{2} \\
z_{3} & z_{4} 
\end{pmatrix};
\quad z_{i} \in \mathbb{C};
\quad u^{\dagger} = u^{-1},
\text{ Det }u=1
\end{equation}
which implies $z_4=z_1^*$ and $z_3=-z_2^*$; that is
\begin{equation}
\label{fourtwo}
u=
\begin{pmatrix}
z_{1} & z_{2} \\
-z_{2}^* & z_{1}^*
\end{pmatrix}
\quad
\text{with}
\quad 
|z_1|^2+|z_2|^2 = 1. 
\end{equation}
\textit{The manifold of the $SU(2)$ group is identical with $S^3$} (this
statement becomes tautological through the equivalence $SU(2) \cong Sp(1)$ or $A_1 = C_1$ in 
Lie algebra notation).

It is instructive to compute now the volume; write $z_1 = |z_1|e^{i\phi}$ and $z_2 = |z_2|e^{i\psi}$ where the
ranges are
\begin{equation}
\label{fourthree}
0 \leq \phi, \psi \leq 2\pi
\end{equation}
and
\begin{equation}
\label{fourfour}
|z_1| = \cos(\beta), \quad |z_1| = \sin(\beta), \quad \text{and} \quad 0 \leq \beta \leq \frac{\pi}{2}
\end{equation}
because $|z_1| > 0$, $|z_2| > 0$, and $|z_1|^2 + |z_2|^2 = 1$. So
\begin{equation}
\label{fourfive}
u=
\begin{pmatrix}
\cos(\beta)e^{i\phi} & \sin(\beta)e^{i\psi} \\
-\sin(\beta)e^{-i\psi} & \cos(\beta)e^{-i\phi}
\end{pmatrix}.
\end{equation}
To calculate the volume, write
\begin{align}
x &= \cos(\beta)\cos(\phi), \quad
y = \cos(\beta)\sin(\phi), \nonumber \\
z &= \sin(\beta)\cos(\psi), \quad
t = \sin(\beta)\sin(\psi) 
\label{eq:foursix}
\end{align}
thus the line element is
\begin{equation}
\label{fourseven}
dx^2 + dy^2 + dz^2 +dt^2 \equiv g(\phi, \psi, \beta)d\phi^2 + \cdots
\end{equation}
with
\begin{align}
dV_{S^3 \subset \mathbb{R}^4} &= \sqrt{\text{Det}\,g}\; d\phi d\psi d\beta \nonumber \\
&= \frac{1}{2}\sin(2\beta)\cdot d\beta \cdot d\phi \cdot d\psi
\label{eq:foureight}
\end{align}
giving of course
\begin{equation}
\label{fournine}
\text{Vol}(S_{R=1}^3) = \frac{1}{2}\; 1\cdot 2\pi \cdot 2\pi = 2\pi^2
\end{equation}
for the ranges given in equations \eqref{fourthree} and \eqref{fourfour}.  We
carried out this elementary calculation because the parameterization was different than 
the one used in equations \eqref{twotwo} and (\ref{eq:twofive}).

A compact group manifold $\EuScript{G}$ has a natural measure, namely the bi-invariant
Haar measure (any locally compact space has a natural left-invariant Haar measure, unique
up to a constant; for compact groups, in which the volume is finite, the name Hurwitz measure
is more appropriate).

The simplest way to measure \textit{invariantly} the volume of $SU(2)$ is to start from the Cartan-Killing metric
in the Lie algebra $L(\EuScript{G})$ of $\EuScript{G}$
\begin{equation}
\label{fourten}
(x, y) = Tr[\text{Ad}(x) \cdot \text{Ad}(y)];
\quad
\text{for }
x,y \in L(\EuScript{G})
\end{equation}
and to induce, by the exponential map $\text{exp} : L(\EuScript{G}) \rightarrow \EuScript{G}$, a Riemannian
metric on $\EuScript{G}$, and hence a finite volume element as $\EuScript{G}$ is always orientable as a manifold
(in particular it is parallelizable) and compact.  One can use any representation  $\Delta : x \rightarrow \Delta(x)$
instead of the adjoint (Ad).

Let us work with the $\EuScript{G}$-invariant metric for $\EuScript{G}=SU(2)$ and see if it coincides with the embedding
metric as $SU(2)=S^3 \subset \mathbb{R}^4$. We start now from the ``Euler'' form in the \textit{defining representation}
\begin{equation}
\label{foureleven}
SU(2) \ni u = e^{i\alpha \sigma_3} e^{i\beta^{\prime} \sigma_2} e^{i\gamma \sigma_3}.
\end{equation}
We need to fix the ranges of the angles $\alpha, \beta^{\prime}$ and $\gamma$; the safest way
is to convert equation \eqref{foureleven} to the form given in equation \eqref{fourfive}, namely
expanding equation \eqref{foureleven} with $\sigma_i^2 =1$
\begin{align}
u &= (\cos(\alpha) + i\sigma_3 \sin(\alpha))(\cos(\beta^{\prime}) + i\sigma_2 \sin(\beta^{\prime}))(\cos(\gamma) + i\sigma_3 \sin(\gamma))
\nonumber \\
&=
\begin{pmatrix}
e^{i(\alpha+\gamma)}\cos(\beta^{\prime}) & e^{i(\alpha-\gamma)}\sin(\beta^{\prime}) \\
-e^{-i(\alpha-\gamma)}\sin(\beta^{\prime} & e^{-i(\alpha+\gamma)}\cos(\beta^{\prime})
\end{pmatrix}
\label{eq:fourtwelve}
\end{align}
and comparing with equation \eqref{fourfive} we obtain
\begin{equation}
\label{fourthirteen}
\beta = \beta^{\prime}, \quad \phi = \alpha+\gamma, \quad \psi = \alpha - \gamma.
\end{equation}
The range of $\{\alpha, \gamma\}$ is \textit{half} of the range $\{\phi, \psi\}$ and
the Jacobian is
\begin{equation}
\label{fourfourteen}
J\biggr(\frac{\phi, \psi}{\alpha, \gamma}\biggl) = 2.
\end{equation}
This is seen also directly from equation (\ref{eq:fourtwelve}):
\begin{equation}
\label{fourfifteen}
u(\alpha, \beta^{\prime}, \gamma) = u(\alpha+\pi, \beta^{\prime}, \gamma+\pi).
\end{equation}
We therefore can choose to write equation \eqref{fourthirteen} as 
\begin{equation}
\label{foursixteen}
0 \leq \beta \leq \frac{\pi}{2}, \quad 0 \leq \alpha \leq \pi, \quad 0 \leq \gamma \leq 2\pi.
\end{equation}

To compute now the volume we proceed as in \cite{MByrdp1,Tilma2}:
\begin{enumerate}
\item{Compute $\frac{\partial{u}}{\partial{\alpha_i}}$ ($1\leq i \leq 3$) in terms of the $\sigma$'s and 
$u$ from equation \eqref{foureleven}; that is, express the holonomic vector fields $\frac{\partial}{\partial \alpha_i}$ ($1\leq i \leq 3$) in terms of the \textit{invariant} anholonomic vector fields $\sigma_i$.}
\item{Invert, to express the invariant frame in terms of the coordinate frame.}
\item{Dualize, to express the invariant co-frame in terms of the one-forms $d\alpha$, $d\beta$ and $d\gamma$.}
\item{Express the volume element as the determinant (Jacobian) of the change of frames.}
\end{enumerate}
This is a standard procedure, it is carried out in detail e.\ g.\ in \cite{Biedenharn}, and the net result is
\begin{equation}
\label{fourseventeen}
dV_{SU(2)} = \sin(2\beta)\cdot d\alpha \cdot d\beta \cdot d\gamma.
\end{equation}
Integration with respect to the ranges given in equation \eqref{foursixteen}, yields
\begin{equation}
\label{foureighteen}
\text{Vol}(SU(2)) = 1 \cdot 2\pi \cdot \pi = 2\pi^2 = V_{embed}(S^3).
\end{equation}
Notice the $\frac{1}{2}$ missing in equation \eqref{fourseventeen} as compared with equation (\ref{eq:foureight})
is compensated with the halving of parameter ranges for $\{\alpha, \beta\}$.
So the \textit{invariant} volume coincides with the \textit{embedding} volume.

Now $SU(2)$ has a center $Z_{2}$, and in fact
\begin{equation}
\label{fournineteen}
SU(2) = Spin(3), \quad SU(2)/Z_2 = SO(3).
\end{equation}
Also $SO(3) \cong \mathbb{R}\mbox{P}^3$, as it is obvious from $SU(2) \cong S^3$.  So let us work the $SO(3)$ case.  

For $\EuScript{G} = SO(3)$ we write
the ``Euler'' formula, namely
\begin{equation}
\label{fourtwenty}
R(\alpha, \beta, \gamma) = R_{o{z}}(\alpha)R_{o{y}}(\beta)R_{o{z}}(\gamma)
\end{equation}
where we are in the \textit{adjoint representation}, e.\ g.\
\begin{equation}
\label{fourtwentyone}
R_{o{z}}(\alpha) = \begin{pmatrix}
\cos(\alpha) & \sin(\alpha) & 0 \\
-\sin(\alpha) & \cos(\alpha) & 0 \\
0 & 0 & 1 
\end{pmatrix}
\end{equation}
etc.  It is well known (e.\ g. \cite{Biedenharn} p.\ 24) that the ranges are
\begin{equation}
\label{fourtwentytwo}
0 \leq \alpha, \gamma \leq 2\pi, \quad 0 \leq \beta \leq \pi.
\end{equation}
We compute the volume as before, starting with $\partial{R}/\partial{\alpha} = (\cdots)R$,
etc.  The final result is
\begin{equation}
\label{fourtwentythree}
dV_{SO(3)} = \sin(\beta)d\alpha \cdot d\beta \cdot d\gamma
\end{equation}
(see e.\ g.\ \cite{Biedenharn} p.\ 58), and the volume is
\begin{equation}
\label{fourtwentyfour}
\text{Vol}(SO(3)) = 2 \cdot 2\pi \cdot 2\pi = 8\pi^2.
\end{equation}
This is the correct volume for $\mathbb{R}\mbox{P}^3$ with geodesic length $2\pi$, which is our case.
From equation \eqref{threethree}, we had obtained $\frac{1}{2}(\text{Vol}(S^3)) = \pi^2$ with
geodesic length equal to $\pi$.  

One might wonder the factor of 2 difference in the angles from equation \eqref{fourseventeen}, namely $\sin(2\beta)$ and here
from equation \eqref{fourtwentythree}, namely $\sin(\beta)$; it is due to the change of generators: with the $\sigma$'s
we have $[\sigma_x, \sigma_y]=2 i \sigma_z$, but with the $J$'s implicit in equation \eqref{fourtwenty} the relation
is the usual one $[J_x, J_y] = i J_z$.  Notice also the Weinstein number $i(\mathbb{R}\mbox{P}^3) = 4$: for equal length
geodesics which is our case, we have $\text{Vol}(\mathbb{R}\mbox{P}^3)/\text{Vol}(S^3) = 8\pi^2/2\pi^2 = 4$.
%---
%---
\section{The Volume of General Groups : $SU(n)$}
\label{sec:five}
A Lie group $\EuScript{G}$ is a symmetric space, $\EuScript{G}=\EuScript{G}_{left} \times \EuScript{G}_{right} / \EuScript{G}_{diag}$, 
and hence of constant scalar curvature; for
rank $>1$ the geodesics depend on directions, in fact they can be dense, as \textit{it is already the case} for
the torus $T^2$.  Another normalization is necessary though.
Let us start with $SU(3)$.  In the vector, or defining representation, $SU(3)$ acts in $\mathbb{C}^{3}$, the 
action being \textit{trans} in the invariant unit sphere $S^5 \subset \mathbb{R}^6 = \mathbb{C}^3$ with the isotropy
subgroup $SU(2)$
\begin{equation}
\label{fiveone}
\frac{SU(3)}{SU(2)} = S^5;\quad SU(2) \rightarrow SU(3) \rightarrow S^5.
\end{equation}
The latter is a principal fibre bundle, in fact, \textit{locally} one certainly has $SU(3) \cong S^3 \times S^5$ but
the \textit{invariant} volume is not quite the product of the volumes of the spheres.  
For many purposes, any compact Lie group can be expressed as a (finitely twisted) topological product
of odd-dimensional spheres.  For a discussion of this point see \cite{Boya}.

Consider the vector \textbf{v}=$\{0,0,1\}$;
it describes the whole of $S^5$ by actions of $SU(3)$; the infinitesimal transformation is $\Bid+\delta\EuScript{G} \equiv \Bid - i\lambda_jdt^j$
where $\lambda_j$ ($1 \leq j \leq 8$) are the Gell-Mann matrices for SU(3) which satisfy
\begin{equation}
\label{fivetwo}
Tr[\lambda_i \lambda_j]=2\delta_{ij}.
\end{equation}

On the other hand, if we describe a point in $S^5$ with locally flat infinitesimal coordinates, we would then have
a Jacobian between the \textit{invariant} coordinates $dt^4 \ldots dt^8$ and the \textit{sphere} coordinates $dx^1 \ldots dx^5$; in fact
\begin{equation}
\label{fivethree}
dV_{inv} = dt^4dt^5dt^6dt^7dt^8 = \frac{\sqrt{3}}{2}\; dx^1dx^2dx^3dx^4dx^5.
\end{equation}
The detailed calculation is in the appendix of \cite{Bernard}.  The factor $\sqrt{3}/2$ is just the
``stretching'' (actually, contracting) of the $\lambda_8$ due to:
\begin{enumerate}
\item{commuting with $\lambda_1, \lambda_2, \lambda_3$ and}
\item{satisfying equation \eqref{fivetwo}.}
\end{enumerate}  
The matrix representation of $\lambda_8$ is necessarily then
\begin{equation}
\label{fivefour}
\lambda_8 = \frac{1}{\sqrt{3}}
\begin{pmatrix}
1 & 0 & 0 \\
0 & 1 & 0 \\
0 & 0 & -2
\end{pmatrix}.
\end{equation}
In other words: for calculating the \textit{invariant volume} it is better to think of the second sphere $S^5$ as being
``stretched'' along a \textit{single} axis by the factor $\sqrt{3}/2$.  Therefore, \textit{with our trace normalization}
(equation \eqref{fivetwo}) the volume of $SU(3)$ is 
\begin{equation}
\label{fivefive}
\text{Vol}(SU(3)) = \frac{\sqrt{3}}{2} \times \text{Vol}(S^5) \times \text{Vol}(S^3) = \frac{\sqrt{3}}{2} \cdot \pi^3 \cdot 2\pi^2 = \sqrt{3}\pi^5
\end{equation}
which is in agreement with \textit{most} of the physics literature, e.\ g. \cite{Marinov2,MByrdp1}.

If we parameterize a generic element of $SU(3)$ as \cite{MByrdp1,MByrdp2,Tilma2}
\begin{equation}
\label{fivesix}
u=e^{i\lambda_3 \alpha_1}e^{i\lambda_2 \alpha_2}e^{i\lambda_3 \alpha_3}e^{i\lambda_5 \alpha_4}
e^{i\lambda_3 \alpha_5}e^{i\lambda_2 \alpha_6}e^{i\lambda_3 \alpha_7}e^{i\lambda_8 \alpha_8},
\end{equation}
one set of ranges of the $\alpha$'s that reproduce equation \eqref{fivefive} is \cite{Tilma2}
\begin{gather}
0 \le \alpha_1,\alpha_5 \le \pi,\quad
0 \le \alpha_2,\alpha_4,\alpha_6 \le \frac{\pi}{2},\nonumber \\
0 \le \alpha_3,\alpha_7 \le 2\pi,\quad
0 \le \alpha_8 \le \sqrt{3}\pi.
\label{eq:fiveseven}
\end{gather}

$SU(3)$ embodies all the complications for the $SU(n)$ series, the calculation of
$\text{Vol}(SU(n))$ in terms of $SU(n-1)$ by induction is now a straightforward matter.

For $SU(n)$ write
\begin{equation}
\label{fiveeight}
\frac{SU(n)}{SU(n-1)} = S^{2n-1},
\quad
SU(n-1) \rightarrow SU(n) \rightarrow S^{2n-1}.
\end{equation}
The ``stretching'' relative to $SU(n-1)$ occurs again \textit{only} in the ``last'' $\lambda_j$
\begin{equation}
\label{fivenine}
\lambda_{last} = \lambda_{n^2-1} = \text{diag}\{1,1,\ldots, 1,-(n-1)\}/\chi
\end{equation}
where $Tr[\lambda_{last}^2] = 2$ implies $\chi = \sqrt{\binom{n}{2}}$ and therefore
\begin{align}
\text{Vol}(SU(n)) &= \frac{\sqrt{\binom{n}{2}}}{n-1} \cdot \text{Vol}(S^{2n-1}) \cdot \text{Vol}(SU(n-1)) \nonumber \\
&= \sqrt{\frac{n}{2(n-1)}}\cdot \frac{2\pi^n}{(n-1)!} \cdot \text{Vol}(SU(n-1)).
\label{eq:fiveten}
\end{align}
An ``invariant'' way of calculating the ``stretching'' is this: write density matrices for $SU(n)$ as
\begin{equation}
\rho = \frac{1}{n}(\Bid_{n} + \boldsymbol{\lambda}\cdot \mathbf{x})
\end{equation}
where $\mathbf{x} \in \mathbb{R}^{n^2-1}$ and the $\lambda$'s satisfy equation \eqref{fivetwo}.  We have $\text{Tr}[\rho]=1$; for
pure states $\rho^2=\rho$ and imposing $\text{Tr}[\rho^2]=1$ we obtain
\begin{equation}
\norm{\mathbf{x}} = x =\sqrt{\binom{n}{2}}.
\end{equation}
So the final formula for $SU(n)$ turns out to be
\begin{align}
\text{Vol}(SU(n))&= \sqrt{\frac{n}{2(n-1)}\; \frac{n-1}{2(n-2)}\cdots \frac{3}{2(2)}}\prod_{k=1}^{n-1}\frac{2\pi^{k+1}}{k!} \nonumber \\
&=\sqrt{\frac{n}{2^{n-1}}}\prod_{k=1}^{n-1}\frac{2\pi^{k+1}}{k!} \nonumber \\
&=\sqrt{n\cdot 2^{n-1}}\;\pi^{(n-1)(n+2)/2}\prod_{k=1}^{n-1}\frac{1}{k!}
\label{eq:fiveeleven}
\end{align}
where we have used $\sum_{k=1}^{n-1}(k+1)=(n-1)(n+2)/2$ in the last step. 
 
Equation (\ref{eq:fiveeleven}) agrees with the corrected volume in \cite{Marinov2} and with
\cite{MacDonald} interpreting his scale $\lambda$ (the Lebesgue measure in $L(SU(3))$) as our
``stretching'' factors; it agrees also with \cite{Bernard}.  
On the contrary, a classical textbook \cite{Gilmore}, and a recent paper \cite{Fuji} omit
these ``stretching'' factors.  

For the full unitary group $U(n)$ there is a \textit{topological} direct product decomposition
\begin{equation}
\label{fivetwelve}
U(n) = SU(n) \times U(1)
\end{equation}
which can be seen, for example, by factorizing a phase in the first vector component:
\begin{equation}
\label{fivethirteen}
U(n) \ni U = \begin{pmatrix}
e^{i\phi} & 0 \\
0 & 1
\end{pmatrix}
u,\quad u\in SU(n).
\end{equation}
The volume of $U(n)$ thus \textit{depends on the radius} of the $U(1)$ factor; if it is $r=1$, i.\ e.\ not 
stretched, then
\begin{equation}
\label{fivefourteen}
\text{Vol}(U(n)) = \sqrt{n\cdot 2^{n+1}}\;\pi^{\binom{n+1}{2}}\prod_{k=1}^{n-1}\frac{1}{k!}.
\end{equation}
Finally, for the projective unitary group $PU(n) = U(n)/U(1) = SU(n)/Z_n$ we obtain, as $\sharp Z_n = n$
\begin{equation}
\label{fivefifteen}
\text{Vol}(PU(n)) = \sqrt{\frac{2^{n-1}}{n}}\pi^{(n-1)(n+1)/2}\prod_{k=1}^{n-1}\frac{1}{k!}.
\end{equation}
%---
%---
\section{Volumes of Other Groups}
\label{sec:six}
For the \textit{orthogonal groups} $O(n)$ and $SO(n)$ we shall proceed in a similar manner;
the rotation group $SO(n)$ acting on the vector representation leaves the unit sphere, $S^{n-1}$, invariant
with the isometry group $SO(n-1)$:
\begin{equation}
\label{sixone}
\frac{SO(n)}{SO(n-1)} = S^{n-1}, \quad SO(n-1) \rightarrow SO(n) \rightarrow S^{n-1}.
\end{equation}
In fact equation \eqref{sixone} is the principal bundle of the tangent to the sphere \cite{Steen}.  There is no 
``stretching'' factor with the Lie algebra convention in the \textit{vector} representation
\begin{equation}
\label{sixtwo}
Tr[\lambda_i \lambda_j] = 2\delta_{ij}
\end{equation}
because the Lie algebra of $SO(n)$ is comprised of antisymmetric matrices, always of the type
\begin{equation}
\label{sixthree}
\lambda_{ij} = 
\begin{pmatrix}
0 & \ldots & \ldots & 0 \\
0 & \ldots & -1 & 0 \\
0 & 1 & \ldots & 0 \\
0 & \ldots & \ldots & 0
\end{pmatrix},
\quad
\text{with }
-Tr[\lambda_{ij}^2] = 2.
\end{equation}
Therefore the volume calculation is elementary, because the spheres
act like those with radius one.

Induction starts at $n=2$.  Of course
\begin{equation}
\label{sixfour}
\text{Vol}(SO(2)) = \text{Vol}(S^1) = 2\pi
\end{equation}
and we have
\begin{align}
\text{Vol}(SO(n)) &= \text{Vol}(S^{n-1})\times \text{Vol}(SO(n-1))\nonumber \\
&= \prod_{d=2}^{n}\text{Vol}(S^{d-1}) \nonumber \\
&= \frac{2^{n-1} \pi^{\frac{(n-1)(n+2)}{4}}}{\prod_{d=2}^{n}\Gamma(\frac{d}{2})}
\; n\geq 2.
\label{eq:sixfive}
\end{align}
which, e.\ g.\, gives $\text{Vol}(SO(3))= 8\pi^2$ which is the same value as in equation \eqref{fourtwentyfour}.
For n even/odd we get
\begin{equation}
\label{sixfiveeven}
\text{Vol}(SO(2n)) =
\frac{2^{n-1}(2\pi)^{n^2}}{\prod_{s=1}^{n-1}(2s)!}
\end{equation}
and
\begin{equation}
\label{sixfiveodd}
\text{Vol}(SO(2n+1)) = \frac{2^n(2\pi)^{n(n+1)}}{\prod_{s=1}^{n-1}(2s+1)!}.
\end{equation}
in agreement with \cite{Gilmore}, and also with \cite{Marinov2} once a
trivial factor of 2 is corrected in the even case.  However we
disagree with \cite{VK}.

The orthogonal group is neither connected nor simply connected; so we have
\begin{equation}
\label{sixsix}
\frac{O(n)}{SO(n)} = Z_{2}, 
\quad
\frac{Spin(n)}{Z_{2}} = SO(n)
\end{equation}
where $Spin(n)$ is the universal double covering of the rotation group ($n\geq 3$).  So we obviously obtain
the result
\begin{equation}
\label{sixseven}
\text{Vol}(O(n))= \text{Vol}(Spin(n)) = 2 \cdot \text{Vol}(SO(n)).
\end{equation}
Notice the first equation implies a topological direct product
$O(n) = SO(n) \times Z_{2} $.  In fact, for odd $n$, this is a direct product of groups.

Notice also that $SO(2n)$ has center $Z_{2}$, therefore the number of central elements in $Spin(2n)$ is $4$, with
two classes:
\begin{align}
\text{Center}(Spin(4n)) &= Z_{2} \times Z_{2}, \nonumber \\  
\text{Center}(Spin(4n+2)) &= Z_{4}.
\label{eq:sixeight}
\end{align}
There are no irreducible \textit{faithful} representations of $Spin(4n)$, so the spin group is represented
through $Spin(4n)/Z_{2}$, which corresponds to $\Delta_L$ and $\Delta_R$, the two chiral \textit{irreps}, and $SO(4n)$
which is the vector representation.  In particular, there are
\textit{three} subgroups in the center of $Spin(4n)$ of type $Z_{2}$.
This explains triality for $SO(8)$, because then $dim\,\Delta_L =
dim\,\Delta_R = dim\,\text{Vector} = 8$.

For the case $Spin(6)$ we have something interesting.  We compute
\begin{equation}
\label{sixnine}
\text{Vol}(Spin(6)) = 2 \cdot \text{Vol}(S^1 \times S^2 \times \cdots \times S^5) = \frac{256}{3}\pi^9
\end{equation}
whereas
\begin{equation}
\label{sixten}
\text{Vol}(SU(4)) = \frac{\sqrt{2}\pi^9}{3}
\end{equation}
even though $SU(4) \cong Spin(6)$!  The ``stretching'' factor is the
culprit, of course.

For the \textit{symplectic groups} $Sp(n)$ the story is pretty much
the same, but now the spheres jump by four: first of all
\begin{equation}
\label{sixeleven}
Sp(1) = SU(2) = Spin(3) =\{\text{Unit Quaternions}\}
\end{equation}
with volume $2\pi^2$ (see equation \eqref{foureighteen}).  Now the induction is based in the fact that
$Sp(n)$ acts in $\mathbb{H}^n$ unitarily, and therefore
\begin{equation}
\label{sixtwelve}
Sp(n) \looparrowright \mathbb{H}^n = \mathbb{C}^{2n} = \mathbb{R}^{4n}; 
\quad Sp(n) \looparrowright S^{4n-1}
\end{equation}
and it is easily seen that the action on the sphere is \textit{trans} with isotropy group
$Sp(n-1)$.  Therefore
\begin{equation}
\label{sixthirteen}
Sp(n-1) \rightarrow Sp(n) \rightarrow S^{4n-1}
\end{equation}
and therefore
\begin{equation}
\label{sixfourteen}
\text{Vol}(Sp(n)) = \text{Vol}(S^3 \times S^7 \times \cdots \times S^{4n-1}).
\end{equation}
It can also been seen that there is no ``stretching'' in the Lie algebra matrices of $Sp(n)$
(\cite{Gilmore}, p.\ 188); the reason is the same as for the orthogonal group.

From equations \eqref{twothree} and \eqref{sixfourteen} we obtain
\begin{align}
\text{Vol}(Sp(n)) &= \prod_{k=1}^n \text{Vol}(S^{4k-1}) = \prod_{k=1}^n \biggr(\frac{2 \pi^{2k}}{(2k-1)!}\biggl) \nonumber \\
&= \frac{2^n \pi ^{2n-1}}{(2n-1)!(2n-3)! \cdots 3!},
\label{eq:sixfifteen}
\end{align}
in full agreement with \cite{Marinov2} and \cite{Gilmore}.  In this
case, the product of spheres $S^3 \times S^7 \times \cdots = Sp(n)$ is
both topological and metric, with radius one spheres.
In particular we obtain $\text{Vol}(Sp(2)) = 2\pi^6/3$ whereas before we obtained
$Spin(5) = 256\pi^6/3$ but $Spin(5) \cong Sp(2)$, corresponding to Cartan's $B_2 = C_2$.
It is remarkable that the \textit{same} normalization, i.\ e.\ $Tr[\lambda_i^2] = 2$ produces such different volumes
in similar groups.  The reason is, of course, that the normalization is performed in 
\textit{different} representations.

As for the \textit{exceptional groups} we just want to add formulas for the two first
cases only, namely $G_2$ and $F_4$.  The groups in the $E$-series, $(E_6, E_7, E_8)$, although fundamental
in $M$-Theory, are yet to be fully understood.  

Now $G_2$ can be \textit{defined} as the automorphism group of the
octonions, or Cayley numbers, $\mathbb{O}$.  The reals $\mathbb{R}
\subset \mathbb{O}$ are of course invariant and so is the norm:
\begin{equation}
\label{sixsixteen}
\text{for } q=hq_0,\; \mathbf{q} \in \mathbb{O}\,\text{ and for }\,\alpha \in \text{Aut}\,\mathbb{O},\; \alpha(q_0)=q_0, \; q_0 \in \mathbb{R};\; \norm{\alpha(q)} = \norm{q}.
\end{equation}
Therefore $G_2$ leaves the set of unit, imaginary, octonions $S^6$ invariant.  One can see also that the action is \textit{trans} with isotropy equal to $SU(3)$ (e.\ g.\ \cite{Rosen}):
\begin{equation}
\label{sixseventeen}
\frac{G_2}{SU(3)} = S^6, \quad SU(3) \rightarrow G_2 \rightarrow S^6.
\end{equation}
So we obtain
\begin{equation}
\label{sixeighteen}
\text{Vol}(G_2) = \sqrt{3}\pi^5 \cdot \frac{16\pi^3}{15} \cdot \xi
\end{equation}
leaving aside a scale factor $\xi$.  

As for $F_4$, it can be \textit{defined} as the isometry group of the Cayley-Moufang plane $\mathbb{O}\mbox{P}^2$; the 
dimension of $F_4$ is 52 and the little group is $Spin(9)$
\begin{equation}
\label{sixnineteen}
\frac{F_4}{Spin(9)} = \mathbb{O}\mbox{P}^2
\end{equation}
as shown in \cite{Baez,Rosen}.  So the volume is 
\begin{equation}
\label{sixtwenty}
\text{Vol}(F_4) = \text{Vol}(\mathbb{O}\mbox{P}^2) \times \text{Vol}(Spin(9)) = \frac{2^{25}\cdot\pi^{28}}{5!\cdot7!\cdot11!}\cdot \xi
\end{equation}
where we have used equations \eqref{threefifteen}, \eqref{sixseven}
and have left a free normalization constant $\xi$.  

We have collected
some of more important volumes in the following table:
\tablehead{ \multicolumn{3}{c}{\textbf{Table \ref{tbl:B}: 
Volumes of Group Manifolds}}\\\hline
Manifold & Normalization & Volume \\\hline \hline}
\begin{center}
\begin{supertabular}{|c|c|l|}
\label{tbl:B}
$SU(n)$ & $Tr[\lambda_i^2]=2$ & equation (\ref{eq:fiveeleven}) \\\hline
$U(n)$ & $Tr[\lambda_i^2]=2$ & equation \eqref{fivefourteen} \\\hline
$SO(2n)$ & \text{vector} & equation \eqref{sixfiveeven} \\\hline
$SO(2n+1)$ & \text{vector} & equation \eqref{sixfiveodd} \\\hline
$Sp(n)$ & \text{fundamental} & equation (\ref{eq:sixfifteen}) \\\hline
\end{supertabular}
\end{center}

%---
%---
\section{Generalized Flag Manifolds}
\label{sec:seven}
States of quantum systems are generally elements of some homogeneous manifold, $X=G/K$.  For example, pure states
lie in $\mathbb{C}\mbox{P}^\infty$, the infinite projective space; if attention is directed to a finite number of independent
states, as it is the case in quantum computing, encryption, entanglement considerations, etc.\ the appropriate frame is
a \textit{finite} dimensional Hilbert space, let us say $\mathbb{C}^{n+1}$.  Pure states here lie in $\mathbb{C}\mbox{P}^n$ which is
equal to $SU(n+1)/U(n)$; marginally mixed states lie within the set of hermitian, unit-trace, positive operators:
\begin{equation}
\label{sevenone}
\{\text{mixed states in $\mathbb{C}^{n+1}$}\} \Leftrightarrow \{\rho \in \mathfrak{E}\, |\, \rho=\rho^\dagger,\, Tr[\rho]=1,\,
\text{Spectrum}[\rho] \geq 0 \},
\end{equation}
where $\mathfrak{E} = \text{End}\mathbb{C}^{n+1}$ are all the
$(n+1)\times(n+1)$ complex matrices.  When the spectrum is
$(1,0,\ldots,0)$ or $\rho^2=\rho$ (idempotency) we recover the \textit{pure} states.

Now the spectrum properties are conserved under conjugation and
therefore the \textit{types} 
of mixed states (including pure ones)
are related to the orbits of the set given in equation \eqref{sevenone} under the
unitary group.  Now, up to 
permutation, the spectral type will 
be indicated by the number of coincident eigenvalues, subject to the
general conditions contained in equation \eqref{sevenone}.  
Permutations are carried out by the Weyl group (the group generated by reflections in hyperplanes defined by the roots, see 
e.\ g.\ \cite{Hel1} (p. 284)).  For $SU(n)$ the group is just $S_n$, the permutation of the $n$ eigenvalues.
Up to permutation, therefore, the spectral types of density matrices are in one-to-one correspondence with partitions
of the number $n$; we shall explicitly show this for the case $n=5$,
the results of which are collected in Table \ref{tbl:C}:
\tablehead{ \multicolumn{5}{c}{\textbf{Table \ref{tbl:C}: 
Spectral Types and Partitions for $n=5$}}\\\hline
Partition & Spectral Type & Orbit & Dimension & States \\\hline \hline}
\begin{center}
\begin{supertabular}{|c|c|c|c|c|}
\label{tbl:C}
$[5]$ & $\lambda_1 =\cdots= \lambda_5=\frac{1}{5}$ &
 $\frac{U(5)}{U(5)}$ & 0 \text{(Single Point)} & Unique (max. entropy)\\\hline
% & & & & (maximal entropy) \\\hline
$[4,1]$ & $\lambda_1 =1, \, \lambda_{i\neq 1} =0$ & $\frac{U(5)}{U(4) \times U(1)}$ & 8 & Pure  $\cong \mathbb{C}\mbox{P}^4$ \\
& $\lambda_i =1-4a, \, \frac{1}{4}>a>0$ & same & 8 & Mix $\cong \mathbb{C}\mbox{P}^4$ \\\hline
$[3,2]$ & $\{a,a,a,b,b\}$ & $\frac{U(5)}{U(3) \times U(2)}$ & 12 & Mixed $\cong Gr_{5,2}$ \\\hline
$[3,1^2]$ & $\{a,a,a,b,c\}$ & $\frac{U(5)}{U(3) \times U(1)^2}$ & 14 & Mixed \\\hline
$[2^2,1]$ & $\{a,a,b,b,c\}$ & $\frac{U(5)}{U(2)^2 \times U(1)}$ & 16 & Mixed \\\hline
$[2,1^3]$ & $\{a,a,b,c,d\}$ & $\frac{U(5)}{U(2) \times U(1)^3}$ & 18 &
Mixed \\\hline
$[1^5]$ & \text{all $\lambda_i$ different} & $\frac{U(5)}{U(1)^5}$ & 20 & Flag manifold \\\hline
% & & & & (generic mixed state) \\\hline
\end{supertabular}
\end{center}
We remind the reader that the number of partitions for large $n$
is only known asymptotically; the spectral type is self-explanatory;
notice pure states lie in $\mathbb{C}\mbox{P}^{n-1}$ ($n=5$ in our case), but
\textit{some} degenerate mixed states will also make a $\mathbb{C}\mbox{P}^{n-1}$
orbit.

The following is to be noticed in the general case $U(n)/K$:
\begin{enumerate}
\item{All these manifolds are homogeneous manifolds, hence they have constant scalar curvature (except $\{\frac{1}{n},\ldots,\frac{1}{n}\}$,
of course, which is just a point).}
\item{Spaces of the type
\begin{equation}
\label{seventhree}
Y=\frac{U(n)}{U(m)\times U(n-m)}
\end{equation}
are called ``Grassmannian.''  Explicitly, $Y$ is the complex Grassmannian of $m$-planes in $\mathbb{C}^n$, to wit
\begin{equation}
\label{sevenfour}
Y=Gr_{n,m}
\end{equation}
(other labels, such as $Gr_{p,q}$ are also used).  In particular
\begin{equation}
\label{sevenfive}
Gr_{n,1} = \mathbb{C}\mbox{P}^{n-1},\quad
Gr_{n,2} = \{\text{bi-planes}\}.
\end{equation}
}
\end{enumerate}
We already used the Grassmannian notation, $Gr$ in section \ref{sec:two}.  
We shall define these orbits of the unitary group in the adjoint
representation as \textit{generalized flag manifolds} following an
extended mathematical usage.  The genuine flag
manifold corresponds to the $[1^n]$ partition:
\begin{equation}
Fl(n) = \frac{U(n)}{U(1)^n}.
\end{equation}
They enjoy interesting mathematical properties; for example, they are
spin and K\"ahler manifolds, in particular symplectic (see for example
\cite{Freed}).  It is remarkable that the simplest space $Fl(3)$
appears as a space of exceptional holonomy \cite{Gibbons}.

We come now to the question of the 
\textit{volumes of the generalized flag manifolds}; because they 
are always \textit{homogeneous} manifolds, $X=G/K$, the volume is, of course,
\begin{equation}
\label{sevensix}
\text{Vol}(X)=\frac{\text{Vol}(G)}{\text{Vol}(K)}
\end{equation}
but, appearances to the contrary, this does not ``cut too much ice'' because the volume of $K \subset G$ depends on 
which subgroup it is identified with!

We illustrate the case $n=3$ ($n=2$ is trivial), before attacking the more general case.  We have
\begin{equation}
\label{sevenseven}
\mathbb{C}\mbox{P}^2 = \frac{U(3)}{U(1)\times U(2)} \quad 
Fl(3) = \frac{U(3)}{U(1)^3} = \frac{U(3)}{U(1)\times U(1) \times U(1)}.
\end{equation}
Now $\mathbb{C}\mbox{P}^2 = SU(3)/U(2)$ also, removing the same $U(1)$ factor, and $U(2)= SU(2)\times U(1)$; this $U(1)$ factor
is ``the long one'' in $\mathbb{C}\mbox{P}^2$, if one remembers that $\mathbb{C}\mbox{P}^2 \cong S^5/S^1$, so $S^1 \subset S^5$ and 
the ``last'' lambda, $\lambda_8$ is of the form diag$\{1,1,-2\}/\sqrt{3}$.  The $U(1)$ group, in this case, has the 
``stretching'' factor $\sqrt{3}/2$ (see section \ref{sec:five}); hence
\begin{equation}
\label{seveneight}
\text{Vol}(\mathbb{C}\mbox{P}^2) = \frac{\text{Vol}(SU(3))}{\text{Vol(SU(2))} \times \text{Vol}(S^1)} = \frac{\sqrt{3}\pi^5}{2\pi^2 \cdot
\frac{\sqrt{3}}{2}\cdot 2\pi} = \frac{\pi^2}{2}
\end{equation}
which coincides with the canonical volume for $\mathbb{C}\mbox{P}^2$ from equation \eqref{threeseven}.  

For the flag manifold we have
\begin{equation}
\label{sevennine}
Fl(3) = \frac{U(3)}{U(1)^3} = \frac{SU(3)}{U(1)\times U(1)} \equiv \frac{SU(3)}{U(1) \times U(1)_{long}}
\end{equation}
thus
\begin{equation}
\label{seventen}
\text{Vol}(Fl(3)) = \frac{\sqrt{3}\pi^5}{2\pi \cdot \frac{\sqrt{3}}{2} \cdot 2\pi} = \frac{\pi^3}{2}
\end{equation}
which coincides, as in the $\mathbb{C}\mbox{P}^2$ case with the ``naive'' calculation without ``stretching''
\begin{equation}
\label{seveneleven}
\text{Vol}(Fl(3))_{naive} = \frac{\text{Vol}(S^1 \times S^3 \times S^5)}{\text{Vol}(S^1 \times S^1 \times S^1)} = \frac{2\pi \cdot 2\pi^2 \cdot 2\pi^3}{2\pi \cdot 2\pi \cdot 2\pi} = \frac{\pi^3}{2}.
\end{equation} 
Now the same the result holds in \textit{all} generality for the generalized flag manifolds!  The
``naive'' calculation (i.\ e.\ neglecting the ``stretching'' factors) \textit{gives the correct results}.
That is
\begin{equation}
\label{seventwelve}
\text{Vol}\biggr(\frac{U(n)}{\underset{\sum q_i = n}\prod U(q_i)}\biggl) = \frac{\text{Vol}(S^1 \times S^3 \times \cdots \times S^{2n-1})}{\prod\text{Vol}(S^1 \times \cdots \times S^{q_i})}.
\end{equation}
The reason is as follows: both the numerator and the 
denominator in equation \eqref{seventwelve} have the same rank!  Therefore, the
dimension of the Cartan subgroups, in fact the 
Cartan subgroups $U(1)_1$, $U(1)_2$, up to $U(1)_{rank}$ are \textit{exactly} the
same in the numerator and the denominator, so the ``stretching'' factors themselves 
cancel completely, and the calculation is thus reduced to the one with the other odd spheres only 
(starting from $S^3$).

So there is nothing else to calculate: for the generic flag manifold the computation is 
\begin{equation}
\label{seventhirteen}
\frac{U(n)}{U(1)^n} = \frac{S^1 \times S^3 \times \cdots \times S^{2n-1}}{S^1 \times S^1 \times \cdots \times S^1} \cong
\mathbb{C}\mbox{P}^1 \times \mathbb{C}\mbox{P}^2 \times \cdots \times \mathbb{C}\mbox{P}^{n-1}
\end{equation}
where
\begin{equation}
\label{sevenfourteen}
\text{Vol}\biggr(\frac{U(n)}{U(1)^n}\biggl) = \prod_{k=1}^{n-1} \frac{\pi^k}{k!}.
\end{equation}
That the generalized flag manifolds have homology of the product of projective spaces times spheres can be proved
easily \cite{Freed} from equation \eqref{seventhirteen}.  

With not too much extra effort we can extend our results to generalized \textit{real} flag manifolds: there are no
``stretching'' factors at all!  We refrain to give explicit formulas, other than to remark that in
\begin{equation}
\label{sevenfifteen}
X = \frac{O(n)}{\underset{\sum n_i = n}\prod O(n_i)}
\end{equation}
the volumes are computed from the volumes of the corresponding orthogonal groups from section \ref{sec:six} with no
corrections.
%---
%---
\section{Final Remarks}
\label{sec:conc}
We hope our normalization conventions are plausible and 
our volume computations useful; as we said in the introduction,
there are many different results written in the literature.  
We have not tried to state the conventions (and, in some cases, the mistakes) 
of all the authors; rather we have attempted to produce a
self-consistent, and uniform way of looking at the volumes of compact
groups and some quotient spaces.

Except for spheres, we don't worry much about parameterizations of
manifolds; for $SU(N)$ see the explicit calculations in \cite{Tilma2,Tilma3}

There are a few points that have been left out, that we want to recall.  We have not attempted to calculate the canonical
volume for the $E_{6,7,8}$ groups, as the defining realizations are rather obscure.  We share the belief \cite{Baez}
that these groups have to be better understood before attempting such a calculation.  Also, the exact \textit{algebraic} characterization
of density matrices (e.\ g.\ equation \eqref{sevenone}) is not done, except in the simplest case of $n=2$, for then
\begin{equation}
\label{eightone}
\text{Vol}(\text{\textit{all} mixed states}) = \int_0^1 r^2 dr (\text{Vol}(S^2 = \mathbb{C}\mbox{P}^1)) = \frac{4}{3}\pi.
\end{equation}
For $n\geq 3$, the generic class is the flag manifold with the matrices congruent to
\begin{equation}
\label{eighttwo}
\text{diag}\;\frac{1}{3}\{1+x_3+\frac{x_8}{\sqrt{3}}, 1-x_3+\frac{x_8}{\sqrt{3}}, 1-\frac{2x_8}{\sqrt{3}}\}
\end{equation}
and the positivity condition is
\begin{equation}
\label{eightthree}
1+x_3+\frac{x_8}{\sqrt{3}}\geq 0,\quad 1-x_3+\frac{x_8}{\sqrt{3}}\geq 0,\quad \frac{\sqrt{3}}{2} \geq x_8
\end{equation}
or, in terms of the minors,
\begin{equation}
\label{eightfour}
\lambda_1 + \lambda_2 + \lambda_3 \geq 0,\quad \lambda_1\lambda_2 + \lambda_2\lambda_3 + \lambda_3\lambda_1 \geq 0,\quad \lambda_1\lambda_2\lambda_3 \geq 0
\end{equation}
which give rise to complicated algebraic restrictions that we shall
not try to pursue further in this paper.

Finally, see the recent paper \cite{Bergman} which attempts to
calculate the volumes of some compact Einstein manifolds which appear
in $M$-theory.
%---
%---
\section{Acknowledgements}
\label{sec:Ack}
Dr. Luis J. Boya's work was sponsored by MCyT grant FPA2000-1252 and wishes to thank Prof. R. F. Schwitters for
his hospitality at C.\ P.\ P.\ (Austin).

%------------------------------------------------------------------------
%------------------------------------------------------------------------

%-------------------------------------------------------------------------
%-------------------------------------------------------------------------

\begin{thebibliography}{99}

\bibitem{tHoft}G. `t Hooft, ``Computation of the quantum effects due to a four-dimensional pseudoparticle,'' 
\textit{Phys. Rev. D}, \textbf{14}, 3432-3450 (1976).

\bibitem{deWitt}B. De Witt, ``Dynamical theories in curved spaces,'' \textit{Rev. Mod. Phys.}, \textbf{29}, 377-397 (1957).

\bibitem{Marinov}M. S. Marinov, ``Invariant volumes of compact groups,'' 
\textit{J. Phys. A: Math. Gen.}, \textbf{13}, 3357-3366 (1980).

\bibitem{Marinov2}M. S. Marinov, ``Correction to `Invariant volumes of compact groups','' 
\textit{J. Phys. A: Math. Gen.}, \textbf{14}, 543-544 (1981).

\bibitem{Marinov3}M. S. Marinov and M. V. Terentyev, ``Dynamics of the group manifold and
path integrals,'' \textit{Fort. d. Phys.}, \textbf{27}, 511-545 (1979).

\bibitem{MacDonald}I. G. MacDonald, ``The volume of a compact Lie group,'' 
\textit{Inven. Math.}, \textbf{56}, 93-95 (1980).

\bibitem{Bernard}C. Bernard, ``Gauge zero modes, instanton determinants, and quantum-chromodynamic calculations,''
\textit{Phys. Rev. D}, \textbf{19}, 3013-3019 (1979).

\bibitem{MByrdp1}M. Byrd, ``Differential geometry on $SU(3)$ with applications to three state systems,''
\textit{J. Math. Phys.}, \textbf{39}, 6125 (1998).

\bibitem{MByrdp2}M. Byrd and E. C. G. Sudarshan, ``$SU(3)$ revisited,''
\textit{J. Phys. A: Math. Gen.}, \textbf{31}, 9255-9268 (1998).

\bibitem{Collab1}L. J. Boya, E. C. G. Sudarshan, and T. Tilma, \textit{in preparation}.

\bibitem{Tilma1}T. Tilma, M. Byrd, and E. C. G. Sudarshan, ``A  parametrization of bipartite systems based on $SU(4)$ Euler angles,'' 
\textit{J. Phys. A: Math. and Gen.}, \textbf{35}, 10445-10465 (2002).

\bibitem{Tilma2}T. Tilma and E. C. G. Sudarshan, ``Generalized Euler angle parametrization for $SU(N)$,''
\textit{J. Phys. A: Math. and Gen.}, \textbf{35}, 10467-10501 (2002).

\bibitem{Wol}J. A. Wolf, \textit{Spaces of Constant Curvature,} Publish or Perish, Berkeley 1977.

\bibitem{Geom1}S. Kobayashi and K. Nomizu, \textit{Foundation of Differential Geometry,} (Chapter IV and XI),
Vols. I and II., J. Wiley and Sons, New York 1963/1969.

\bibitem{Hel1}S. Helgason, \textit{Differential Geometry, Lie Groups, and Symmetric Spaces,}
(Chapter VIII, Section 10), Am. Math. Soc., Providence, RI 2001.

\bibitem{Santan} In $\mathbb{C}\mbox{P}^2$ for example, we have a complex structure, hence real bi-planes become
complex lines: there are ``holomorphic'' complex lines $\cong \mathbb{C}\mbox{P}^1 \cong S^2$ with sectional curvature
$K=1$, and orthogonal complex lines $\cong \mathbb{R}\mbox{P}^2$ with $K=\frac{1}{4}$.  Interpolating complex lines have
intermediate curvatures.  LJB thanks M. Santander (Valladolid) for elucidating this point. 

\bibitem{Besse1}A. L. Besse, \textit{Manifolds All of Whose Geodesics are Closed,} (Chapter III),
Springer, Berlin 1978.

\bibitem{Wei1}A. Weinstein, ``On the volume of manifolds all of whose geodesics are closed,'' 
\textit{J. Diff. Geom.}, \textbf{29}, 29-41 (1974).

\bibitem{Gib}G. W. Gibbons and C. N. Pope, ``$\mathbb{C}\mbox{P}^2$ as a gravitational instanton,''
\textit{Comm. Math. Phys.}, \textbf{61}, 239-248 (1978).

\bibitem{Steen}N. F. Steenrod, \textit{The Topology of Fibre Bundles,} Princenton, New Jersey 1951.

\bibitem{Baez}J. C. Baez, ``The octonions,'' \textit{Bull. Am. Math. Soc},
\textbf{39}, 145-206 (2002).

\bibitem{Biedenharn}L. C. Biedenharn and J. D. Louck, \textit{Angular Momentum in 
Quantum Physics : Theory and Application,} in 
\textit{Encyclopedia of Mathematics and its Applications: Vol. 8,}
ed.\ by Gian-Carlo Rota, Addison-Wesley, Massachusetts 1981.

\bibitem{Boya}L. J. Boya, ``The geometry of compact Lie groups,'' \textit{Rep. Math. Phys.},
\textbf{30}, 149-167 (1991).

\bibitem{Gilmore}R. Gilmore, \textit{Lie Groups and Lie Algebras,} (Chapter V.7),
J. Wiley and Sons, New York 1974.

\bibitem{Fuji}K. Fuji, ``Introduction to Grassmannian manifolds and quantum computations,'' LANL eprint
\texttt{quant-ph$\backslash$0103011} (2001).

\bibitem{VK}N. Ja. Vilenkin and A. U. Klimyk, \textit{Representation of Lie Groups
and Special Functions: Vol. 2,} Kluwer Academic Publishers, Netherlands 1993.

\bibitem{Rosen}L. Rosenthal, \textit{Geometry of Lie Groups,} Kluwer Academic, Amsterdam 1990.

\bibitem{Freed}D. Freed, ``Flag manifolds and infinite dimensional K\"ahler geometry,'' in  \textit{Infinite Dimensional Groups}
ed. by V. Kac, Springer, Berlin 1985 and \textit{personal communication}.

\bibitem{Gibbons}M. Cveti\~c, G. W. Gibbons, H. Lu, and C. N. Pope, 
``Cohomogeneity one manifolds of $Spin(7)$ and $G(2)$ holonomy,''
\textit{Phys. Rev. D}, \textbf{65}, 106004 (2002).

\bibitem{Tilma3}T. Tilma and E. C. G. Sudarshan, ``Generalized Euler
angle parameterization for $U(N)$ with applications to $SU(N)$ coset 
volume measures,'' LANL eprint \texttt{math-ph$\backslash$0210057} (2002).

\bibitem{Bergman}A. Bergman and C. P. Herzog, ``The volume of some non-spherical horizons and the AdS/CFT Correspondence,''
LANL eprint \texttt{hep-th$\backslash$0108020} (2001).

\end{thebibliography}
\end{document}